\documentclass[aps,prd,preprint,groupedaddress,showpacs]{revtex4-1}
\usepackage{amsmath,amssymb}
\usepackage{graphicx}
\begin{document}
\title{Initial directional singularity in inflationary
models} \author{L. Fern\'andez-Jambrina}
\email[]{leonardo.fernandez@upm.es}
\homepage[]{http://dcain.etsin.upm.es/ilfj.htm}
\affiliation{Matem\'atica Aplicada, E.T.S.I. Navales, Universidad
Polit\'ecnica de Madrid,\\
Arco de la Victoria 4, \\ E-28040 Madrid, Spain}%
\date{\today}
\begin{abstract}
In \cite{haro} a new cosmological model is proposed with no big bang
singularity in the past, though past geodesically incomplete.  This
model starts with an inflationary era, follows with a stiff matter
dominated period and evolves to accelerated expansion in an
asymptotically de Sitter regime in a realistic fashion.  The big bang
singularity is replaced by a directional singularity.  This
singularity cannot be reached by comoving observers, since it would
take them an infinite proper time lapse to go back to it.  On the
contrary, observers with nonzero linear momentum have the singularity
at finite proper time in their past, though arbitrarily large.  Hence,
the time lapse from the initial singularity can be as long as desired,
even infinity, depending on the linear momentum of the observer.  This
conclusion applies to similar inflationary models.  Due to the
interest of these models, we address here the properties of such
singularities.
\end{abstract}

\maketitle

\section{Introduction}

The accelerated expansion of our universe \cite{snpion,
Davis:2007na,WoodVasey:2007jb,Leibundgut:2004,wmap} has motivated the
consideration of either new ingredients in the energy content of
cosmological models
\cite{Padmanabhan:2006ag,Albrecht:2006um,Sahni:2006pa} or corrections
to the general theory of gravitation compatible with observations 
\cite{Maartens:2007,Durrer:2007re,Padmanabhan:2007xy,modigravi}.

As a consequence, some energy conditions are violated by these new
ingredients with the result of new future scenarios for our universe
in the form of new singularities (big rip, sudden singularities\ldots)
or nonsingular asymptotic behaviors observationally undistinguishable from
singularities (pseudorip, little rip\ldots). Some of these 
singularities are weak in the sense that the universe can be extended 
beyond the singularity and in consequence it cannot be considered 
 the end of the universe. These phenomena have 
also been discovered recently in inflationary models \cite{graham}.

But these singular behaviors may also appear at the beginning of our 
universe, replacing the traditional big bang as initial singularity. 
One of these models is \cite{haro}, but other inflationary models 
\cite{borde} follow a similar pattern.

The model \cite{haro} proposes a simple equation of state which 
succeeds in removing the big bang singularity, replacing it by 
another one, dubbed little bang in analogy with the little rip, and 
producing an inflationary era. A phase transition stops the inflation 
until in the far future accelerated expansion is dominant. An 
interesting feature shown in \cite{haro} is that the new singularity is at 
infinite cosmic time, for comoving observers, but at finite proper 
time for noncomoving observers. This resembles the behavior of 
directional singularities in \cite{hidden,grandrip}.

We would like to comment here the nature and properties of these
initial singularities appearing in some inflationary cosmological
models.
We begin by reviewing the possible singular scenarios in Section~2,
paying special attention to directional singularities in Section~3 in
order to frame the inflationary model in Section~4.  The derived
conclusions are summarized in the final Section.

\section{Cosmological singularities}

In \cite{grandrip} a thorough classification of cosmological
singularities has been provided both at finite and infinite coordinate
time, obtained in terms of either the behavior of the barotropic index
$w$ for flat models of scale factor $a$ or equivalently the
deceleration parameter $q$,
\[w=
\frac{p}{\rho}=-\frac{1}{3}-\frac{2}{3}\frac{a\ddot a}{\dot 
a^2},\qquad q=-\frac{a\ddot a}{\dot a^2}=\frac{1+3w}{2},\]
where $\rho$ is the energy density and $p$ is the pressure of the 
model and the dot stands for derivative with respect to coordinate 
time $t$.

This classification makes use of generalized power expansions 
\cite{visser} in coordinate time of the deviation $h$ from the pure
cosmological constant case,
\[ w(t)=-1+\frac{2}{3}h(t),\qquad q(t)=-1-h(t),\] and extends the one in 
\cite{Nojiri:2005sx}, which has been enlarged in \cite{IV, yurov, sesto}. We shall not include here nonsingular 
behaviors such as little rip \cite{little}, pseudorip  \cite{pseudo} 
and the  little sibling of the big rip \cite{sibling}, since we are 
concerned just with singularities, though they are also taken into 
account in \cite{grandrip}. The classification can be summarized 
as follows:

\begin{itemize}    
   \item Type -1: ``Grand bang/rip'': \cite{grandrip} The scale factor
   vanishes or blows up at $w=-1$.  The Hubble ratio, the energy
   density and the pressure blow up. These are strong singularities.
   
   \item Type 0: ``Big bang'': The scale factor vanishes at
   $w\neq-1$.  The Hubble ratio, the energy density and the pressure blow
   up. These are also strong.
   
   \item Type I: ``Big rip'' \cite{Caldwell:2003vq}: The scale 
   factor, the Hubble ratio, the energy density and the
   pressure blow up.  Null geodesics are complete, but not 
   timelike geodesics. They are strong singularities. 

   \item Type II: ``Sudden singularities'' \cite{sudden,suddenfirst}: 
   They have been also dubbed ``quiescent singularities''
   \cite{quiescent}: The scale factor, the Hubble ratio and the
   energy density remain finite, whereas the pressure blows up.  That
   is, the second derivative of the scale factor diverges.  Some
   subcases have been dubbed big brake \cite{brake} and big boost
   \cite{boost}.   These are weak singularities 
   \cite{suddenferlaz} and the models just violate the dominant 
   energy condition.

   \item Type III: ``Big freeze'' \cite{freeze} or ``finite scale factor 
   singularities'': The scale factor remains finite, but the Hubble 
   factor, the energy density and the pressure blow up. That is, the 
   first derivative of the scale factor is singular. Depending on the 
   definition used \cite{tipler,krolak}, they can be either strong or weak \cite{puiseux}.

   \item Type IV \cite{tsagas}: The scale factor, the Hubble ratio,
   the energy density and the pressure are finite, whereas higher
   derivatives of the scale factor blow up.  They are dubbed
   ``generalized sudden singularities'' if the barotropic
   index $w$ is finite \cite{sesto} and big separation if it blows up 
   with vanishing pressure and energy density. These are weak 
   singularities.
   
   \item Type V: ``$w$-singularities'' \cite{wsing, loitering}: The scale factor, the Hubble ratio,
   the energy density, the pressure and higher derivatives of the 
   scale factor are finite, whereas the barotropic index 
   $w$ blows up. They are weak singularities \cite{barotrope}.

\item Type $\infty$: ``Directional singularities'' \cite{hidden}: These type of
singularities appear at infinite coordinate time, but at finite proper
time, at least for some observers.  In this sense they are
directional.  These are p.p. curvature singularities (curvature
singularities along a parallelly transported basis) \cite{HE}.  We pay
a little attention to this overlooked type of singularities.
\end{itemize}

This analysis has been done at classical level.  It must be taken into
account that some of these singularities have been shown to be
removable on considering quantum gravity \cite{carlson} and loop
quantum gravity corrections \cite{vidotto}.

\section{Type $\infty$ singularities}

Type $\infty$ singularities appear at coordinate time $t=\pm \infty$. 
In general, this time is inaccessible, but this is 
not so in certain cosmological models. 

For a flat FLRW cosmological model with scale factor $a(t)$ and metric
\begin{equation}ds^2=-dt^2+a^2(t)\left(dr^2+ r^2\left(d\theta^2+\sin^2\theta
d\phi^2\right)\right),\label{metric}\end{equation}
we notice \cite{hidden} that the system of equations for geodesic curves, 
followed by nonaccelerated observers ($\delta=1$) and lightlike 
particles ($\delta=0$) with specific linear momentum $P$, can be 
reduced to
\begin{subequations}\begin{eqnarray}\label{geods}
     \frac{dt}{d\tau}&=&\sqrt{\delta 
     +\frac{P^2}{a^2(t)}},\label{geods1}\\\frac{dr
    }{d\tau}&=&\pm\frac 
    {P}{a^2(t)}.\label{geods2}\end{eqnarray}\end{subequations}
for constant $\theta$ and $\phi$, due to the symmetry of these 
models, and where $\tau$ is the intrinsic or proper time as measured by the 
observer. 

For null geodesics we have
\[\frac{dt}{d\tau}=\frac{P}{a(t)} \Rightarrow 
\Delta\tau=\frac{1}{P}\int_{-\infty}^ta(t)\,dt.\]

Hence, for the initial \emph{event} $t=-\infty$ to be at a finite 
proper time lapse $\Delta\tau$ of an event at $t$, we require
\begin{equation} \label{condit}
    \int_{-\infty}^ta(t)\,dt<\infty.
    \end{equation}

That is, singular behavior at $t=-\infty$ only may appear if the 
scale factor is an integrable function of coordinate time. This means 
that it is necessary, but not sufficient, that $a(t)$ tends to zero when $t$ tends to $-\infty$.

Similarly, for timelike geodesics with nonzero $P$,
\[\Delta \tau=\int_{-\infty}^t\frac{dt}{\sqrt{1+\frac{P^2}{a^2(t)}}}<\frac{1}{P}\int_{-\infty}^t
a(t)\,dt,\] the proper time lapse to $t=-\infty$ is finite if the time 
lapse for lightlike geodesics is finite and then $t=-\infty$ is 
accessible for these observers.

Hence, condition (\ref{condit}) implies that both lightlike and
timelike geodesics with nonzero $P$ have $t=-\infty$ at a finite
proper time lapse in their past.

On the contrary, comoving observers, following timelike geodesics 
with $P=0$, have $d\tau=dt$ and therefore $t=-\infty$ is for them at 
an infinite proper time lapse in the past and cannot have experienced 
the singularity.

This is the reason why Type $\infty$ singularities are directional, 
in the sense that they are accessible for causal geodesics, except for 
those with $P=0$.

According to \cite{grandrip}, Type $\infty$ singularities may appear 
in three instances:

\begin{itemize}
    \item  Finite $\int_{-\infty} h\,dt$, $h(t)>0$: $a_{-\infty}=0$, 
    $\rho_{-\infty}=\infty$, $p_{-\infty}=-\infty$, $w_{-\infty}=-1$. 
    They differ from little rip in the sign of $h(t)$, so they can be 
    dubbed \emph{little bang} if it is an initial singularity or 
    \emph{little crunch} \cite{grandrip} if it is a final singularity. Instances of 
    this case are models with scale factor $a(t)\propto e^{-\alpha(-t)^p}$ 
    with $p>1$, $\alpha>0$.

    \item $h_{-\infty}=0$, $|h(t)|\gtrsim |t|^{-1}$, $h(t)<0$:
    $a_{-\infty}=0$, $\rho_{-\infty}=0$, $p_{-\infty}=0$,
    $w_{-\infty}=-1$.  By changing the sign of $h(t)$ we obtain a sort
    of little rip with vanishing asymptotic energy density and
    pressure.  Examples for this case are models with scale factor
    $a(t)\propto e^{-\alpha(-t)^p}$ with $p\in(0,1)$, $\alpha>0$.

    \item  Finite $h_{-\infty}\in(-1,0)$: $a_{-\infty}=0$, $\rho_{-\infty}=0$, 
    $p_{-\infty}=0$, finite $w_{-\infty}\neq-1$. This is the case, 
    for instance, of models with $a(t)\propto t^{-p}$, $p>1$, as the 
    ones studied in \cite{hidden}.
\end{itemize}

It is interesting to check the strength of these singularities in 
order to know if the model can be extended beyond the singularity.

There are several definitions of strong singularities. The concept 
comes up first in \cite{ellis} by defining a strong curvature 
singularity as one for which no object ``can arrive intact at the 
singularity''. 

Tipler \cite{tipler} clarifies the concept by defining 
a strong curvature singularity as one for which ``any object hitting 
it is crushed to zero volume''. The volume of the object is 
rigorously defined by any three linearly independent spacelike 
vorticity-free Jacobi fields orthogonal to the velocity of the 
geodesic. This definition is equivalent to inextendibility of the 
spacetime in a continuous fashion beyond the singularity.

In the context of cosmic censorship Kr\'olak \cite{krolak} proposed
another definition which requires that, instead of a vanishing volume
of the \emph{object}, the derivative of the volume must be negative
close to the singularity.

Such definitions are complex to apply from scratch, but fortunately
there are necessary and sufficient conditions for their requirements
\cite{clarke}. They are even simpler in our case, since FLRW 
spacetimes are conformally flat. 

For instance, according to Tipler, a null geodesic ends up at a strong
singularity at proper time $\tau_{0}$ if and only if
\begin{equation}\label{suftipler}
 \int_{0}^{\tau}d\tau'\int_{0}^{\tau'}d\tau''R_{ij}u^{i}u^j
\end{equation}
blows up as $\tau$ tends to $\tau_{0}$. $R$ is the Ricci tensor 
of the spacetime and $u$ is the velocity of the geodesic. 

According to Kr\'olak, a null geodesic ends up at a strong
singularity at $\tau_{0}$ if and only if
\begin{equation}\label{sufkrolak}
 \int_{0}^{\tau}d\tau'R_{ij}u^{i}u^j
 \end{equation}
blows up as $\tau$ tends to $\tau_{0}$.

For timelike geodesics the previous conditions become just sufficient 
conditions.

Let us check these requirements for the first two subtypes of 
singularities. For the third subtype the strength was checked in 
\cite{hidden}. 

For a null geodesic, the components of the velocity $u$ are
\[u^t= \frac{dt}{d\tau}=\frac{P}{a},\qquad u^r=\frac{dr}{d\tau}=\pm\frac{P}{fa^2},\]
and hence the Ricci curvature along the geodesic takes the expression
\begin{equation}\label{ricci}
R_{ij}u^iu^j\,d\tau
=2P^2\left(\frac{\dot a^2}{a^4}-\frac{\ddot a}{a^3}\right)d\tau
=2P\left(\frac{\dot a^2}{a^3}-\frac{\ddot a}{a^4}\right)dt
=-2P\ddot xe^{-x}\,dt,
\end{equation}
 in terms of $x(t)=\ln a(t)$.
%


For timelike geodesics, 
\[u^t=\sqrt{1+\frac{P^2}{a^2}},\qquad u^r=\pm \frac{ P}{a^2},\] 
since $a_{-\infty}=0$ we have
\[R_{ij}u^i u^j\,d\tau=
\frac{-\frac{3\ddot a}{a}+2P^2\left(\frac{\dot a^2}{a^4}-\frac{\ddot a}{a^3}
     \right)}{\sqrt{1+\frac{P^2}{a^2}}}dt\simeq\left(
-\frac{3\ddot a}{P}+2P\left(\frac{\dot a^2}{a^3}-\frac{\ddot a}{a^2} 
\right)\right)dt.\]

The second term already appears for null geodesics.  The first term is
smaller than the second one, since $w\simeq -1$ for these models.
Therefore the conclusions for null geodesics are valid also for
timelike geodesics close to these directional singularities.

In order to have finite integrals of (\ref{ricci}) it is necessary 
that $\ddot x$ tends to zero when $t$ tends to $-\infty$, since 
$e^{-x}=a^{-1}$ tends to infinity for directional singularities, and 
hence $x$ tends to $-\infty$ either. 

The function $x$ should be then a divergent function 
of time with decreasing concavity $\ddot x$, asymptotically tending to zero. 
This happens with functions which behave asymptotically as $x(t)\simeq
-(-t)^{p}$, with $0<p<2$.  Faster diverging functions have nonzero
asymptotic acceleration and functions decreasing more slowly do not
diverge at infinity.
%

These sort of functions produce divergent integrals of the Ricci 
curvature and hence we are to conclude that all Type $\infty$ singularities are 
strong according to Tipler's and Kr\'olak's criteria.

\section{The model}

The model proposed in \cite{haro} has a scale factor of the form
\begin{eqnarray}
a(t)=\left\{\begin{array}{ccc}
a_Ee^{-\frac{1}{6\gamma}\left( 1+\frac{2H_f}{H_e}+\sqrt{\frac{8H_f}{H_e}}   
\right)\left[e^{-{3\gamma}H_e\,t}-1  \right]} 
e^{\frac{H_e}{2}t}&\mbox{if}& t<0
\\
a_E\left(\frac{3\gamma}{2}( H_e+\sqrt{2H_eH_f}   )t+1 
\right)^{\frac{2}{3\gamma}}e^{{H_f}t}&\mbox{if}& t\geq 0,
\end{array}
\right.
\end{eqnarray}
where $a_{E}$, $\gamma$, $H_{e}, H_{f}, H_{E}$ are parameters of the model.

Taking into account the values of these parameters in the model, the scale factor 
can be approximated as
\begin{eqnarray}
a(t)\simeq\left\{\begin{array}{ccc}
a_Ee^{-\frac{1}{6\gamma}\left[e^{-{3\gamma}H_e\,t}-1  \right]}e^{\frac{H_e}{2}t}&\mbox{for}& t<0
\\
a_E\left(\frac{3\gamma}{2}H_e\,t+1 \right)^{\frac{2}{3\gamma}}e^{{H_f}t}&\mbox{for}& t\geq 0.
\end{array}
\right.
\end{eqnarray}

We are interested in the behavior of the model for very small negative $t$. For that 
era, the barotropic index of the model is\[
w(t)\simeq -1+\frac{2}{18\gamma}e^{3\gamma H_et},\qquad 
h(t)\simeq\frac{e^{3\gamma H_et}}{6\gamma}.\]

In \cite{haro} it is shown that this model has no big bang singularity
and there is no initial singularity in cosmic time.  However, a
singularity appears at finite proper time in the past for noncomoving
observers.

This can be derived within our formalism for this model and similar
ones, since in this case it is clear that $h(t)$ is an integrable
function of coordinate time and therefore the model has a Type
$\infty$ singularity of the first kind in our classification
($a_{\infty}=0$, $\rho_{\infty}=\infty$, $p_{\infty}=-\infty$,
$w_{\infty}=-1$).



\section{Concluding remarks}

We have shown that the model in \cite{haro} and similar inflationary 
models \cite{borde} with the property 
\[\int_{-\infty}^Ta(t)\,dt<\infty,\] for some time $T$ have a directional 
singularity as initial singularity, which is accessible in finite
proper time only for null geodesics and timelike geodesics with finite
linear momentum $P$.  Comoving observers, following cosmological fluid
worldlines, have not experienced the initial singularity, since it would
have taken them infinite proper time to reach present time.  Their
geodesic trajectories are complete towards the past.  

This does not happen in other cosmological models for which there is
no such discrepancy between the finiteness of proper time and
coordinate time lapses.  

The absence of a big bang singularity is an interesting feature for a
cosmological model, even though the curvature still blows up at the
new singularity.  Milder singularities with vanishing, instead
of diverging, energy density and pressure could be obtained with 
similar models, but with nonintegrable $h(t)$.


For a model starting with a big bang singularity, the proper time of
comoving observers is finite and defines the maximum age of the Universe 
that can be experienced by nonaccelerated observers. 

On the contrary, for a model with a little bang singularity, the age
of the universe in the previous sense is infinite and the proper time
as measured by nonaccelerated observers can be as large as desired by
diminishing their linear momentum $P$.


It is an intriguing feature the idea of initial singularity in these
models, with observers for which the universe extends indefinitely to
the past, avoiding the singularity.  However, as it has been pointed
in Section II, this is a pure classical analysis.  It is expected that
the necessary quantum effects to be considered on approaching the
singularities may appease them as it has happened in other instances.

    

\begin{thebibliography}{99}
    
\bibitem{haro} J. Haro, J. Amor\'os, S. Pan, Phys.  Rev.  D
\textbf{93}, 084018 (2016).

\bibitem{snpion}
A. G. Riess et al. [Supernova Search Team Collaboration],
Astron. J. {\bf 116} (1998)  1009  [arXiv:astro-ph/9805201];
S. Perlmutter et al. [Supernova Cosmology Project Collaboration],
Astrophys. J. 517, 565 (1999)
[arXiv:astro-ph/9812133].
\bibitem{Davis:2007na}
 T.~M.~Davis {\it et al.},
 Astrophys.\ J.\  {\bf 666} (2007) 716
 [arXiv:astro-ph/0701510].
\bibitem{WoodVasey:2007jb}
 W.~M.~Wood-Vasey {\it et al.}  [ESSENCE Collaboration],
 Astrophys.\ J.\  {\bf 666} (2007) 694
 [arXiv:astro-ph/0701041].
\bibitem{Leibundgut:2004} B. \,Leibundgut, 
in Reviews of Modern Astronomy  {\bf 17} (2004)
edited by R. E. Schielicke (Wiley-VCH, Weinheim)
\bibitem{wmap}  D.~N.~Spergel {\it et al.}  [WMAP Collaboration],
 Astrophys.\ J.\ Suppl.\  {\bf 148} (2003) 175
 [arXiv:astro-ph/0302209];
 D.~N.~Spergel {\it et al.}  [WMAP Collaboration],
 Astrophys.\ J.\ Suppl.\  {\bf 170} (2007) 377
 [arXiv:astro-ph/0603449];
 J.~Dunkley {\it et al.}  [WMAP Collaboration],
Observations:
 arXiv:0803.0586 [astro-ph], E.~Komatsu {\it et al.}  [WMAP Collaboration],
arXiv:0803.0547 [astro-ph].
\bibitem{Padmanabhan:2006ag}
T.~Padmanabhan,
 AIP Conf.\ Proc.\  {\bf 861} (2006) 179
 [arXiv:astro-ph/0603114].
\bibitem{Albrecht:2006um}
 A.~Albrecht {\it et al.},
 arXiv:astro-ph/0609591.
\bibitem{Sahni:2006pa}
 V.~Sahni and A.~Starobinsky,
 Int.\ J.\ Mod.\ Phys.\  D {\bf 15} (2006) 2105
 [arXiv:astro-ph/0610026].
\bibitem{Maartens:2007} Roy Maartens, J. Phys.: Conf. Ser. 68 (2007)
012046.
\bibitem{Durrer:2007re}
 R.~Durrer and R.~Maartens,
 Gen.\ Rel.\ Grav.\  {\bf 40} (2008) 301
 [arXiv:0711.0077 [astro-ph]].
\bibitem{Padmanabhan:2007xy}
 T.~Padmanabhan,
 arXiv:0705.2533 [gr-qc].
 \bibitem{modigravi}L.~Fern\'andez-Jambrina, R. Lazkoz, \emph{Phys. Lett. 
B} \textbf{670}, 254-258 (2009) [arXiv:0805.2284].



\bibitem{graham} J.D. Barrow, A.A.H. Graham, \emph{Phys.  Rev.  D}
\textbf{91}, 083513 (2015).

\bibitem{borde} A. Borde, A.H. Guth and A. Vilenkin, \emph{Phys. Rev. Lett.} \textbf{90}, 
151301 (2003).


\bibitem{hidden}L.~Fern\'andez-Jambrina, \emph{
Phys.\ Lett.\ B} \textbf{656}, 9 (2007) [arXiv:gr-qc/0704.3936].


\bibitem{grandrip} L. Fern\'andez-Jambrina, Phys.  Rev.  D
\textbf{90}, 064014 (2014).

\bibitem{visser} C. Catto\"en and  M. Visser,
Class.\ Quant.\ Grav.\  {\bf 22} (2005) 4913
[arXiv:gr-qc/0508045].

\bibitem{Nojiri:2005sx}
 S.~Nojiri, S.~D.~Odintsov and S.~Tsujikawa,
 Phys.\ Rev.\  D {\bf 71} (2005) 063004

\bibitem{IV} M.P. D\c abrowski, K. Marosek, \emph{JCAP} \textbf{2013} 
02, 012 (2013).

\bibitem{yurov} A.V. Yurov, \emph{Phys. Lett. B} \textbf{689}, 1 
(2010).

\bibitem{sesto} M.P. D\c abrowski, K. Marosek, A. Balcerzak, 
\emph{Memorie della Societa Astronomica
Italiana} \textbf{85}, 44-49 (2014) [arxiv:1308.5462].

\bibitem{little} P.H. Frampton, K.J. Ludwick, R.J. Scherrer, \emph{Phys.
Rev.  D} \textbf{84}, 063003 (2011); P.H. Frampton, K.J. 
Ludwick, S. Nojiri, S.D. Odintsov, R.J. Scherrer, \textbf{Phys. Lett. 
B} \textbf{708}, 204 (2012).

\bibitem{pseudo} P.H. Frampton, K.J. Ludwick, R.J. Scherrer,
\emph{Phys.  Rev.  D} \textbf{85}, 083001 (2012)

\bibitem{sibling} M. Bouhmadi-Lopez, A. Errahmani, P. Martin-Moruno, 
T. Ouali, Y. Tavakoli, arXiv:1407.2446 (2014).

\bibitem{Caldwell:2003vq}
 R.~R.~Caldwell, M.~Kamionkowski and N.~N.~Weinberg,
 Phys.\ Rev.\ Lett.\  {\bf 91} (2003) 071301
 [arXiv:astro-ph/0302506].

\bibitem{sudden} J.D. Barrow, \textit{Class. Quant. Grav.} {\bf 21}, L79 (2004)
;  S. Nojiri, S.D. Odintsov, \textit{Phys.\ Lett. B} 
 {\bf 595}, 1 (2004); 
J.D. Barrow,
  \textit{Class.\ Quant.\ Grav.}\  {\bf 21}, 5619 (2004); 
K. Lake,
  \textit{Class.\ Quant.\ Grav.}  {\bf 21}, L129 (2004);  
 S. Nojiri, S.D. Odintsov, \textit{Phys.\ 
  Rev.\ D} {\bf 70}, 103522 (2004); 
  M.P. D\c abrowski,
  \textit{Phys.\ Rev.\ D} {\bf 71}, 103505 (2005); 
L.P. Chimento,  R. Lazkoz,
\textit{Mod.\ Phys.\ Lett.\ A} {\bf 19}, 2479 (2004) ; 
 M.P. D\c abrowski,
\textit{Phys.\ Lett.\ B} {\bf 625}, 184 (2005); 
J.D. Barrow, A.B. Batista, J.C. Fabris, S. Houndjo, \emph{Phys. Rev. 
D} \textbf{78}, 123508 (2008); J.D. Barrow, S.Z.W. Lip, \emph{Phys. 
Rev. D} \textbf{80}, 043518 (2009); S. Nojiri, S.D. Odintsov,
\emph{Phys. Rev. D} \textbf{78}, 046006 (2008); 
J.D. Barrow, S. Cotsakis, A. Tsokaros, \emph{Class. Quant. Grav.} 
\textbf{27}, 165017 (2010); 
J.D. Barrow, S. Cotsakis, A. Tsokaros, [arXiv:1003.1027] (2010); 
P. Singh, \emph{Phys. Rev. D} \textbf{85}, 104011 (2012); 
T. Denkiewicz, M.P. D\c abrowski, H. Ghodsi, M.A. Hendry,
\emph{Phys. Rev. D} \textbf{85}, 083527 (2012).

\bibitem{suddenfirst} J.D. Barrow, G.J. Galloway, F.J. Tipler, 
\emph{MNRAS} \textbf{223}, 835 (1986).

\bibitem{quiescent} Y. Shtanov, V. Sahni, \emph{Class. Quant. Grav.}
 \textbf{19}, L101 (2002) [arXiv:gr-qc/0204040].


\bibitem{brake} V. Gorini, A.Y. Kamenshchik, U. Moschella, 
V. Pasquier,
 \emph{PRD} {\bf 69}, 123512 (2004).
 
\bibitem{boost} A.O. Barvinsky, C. Deffayet, A.Yu. Kamenshchik, 
\emph{JCAP} \textbf{05}, 034 (2010) [arxiv:0801.2063].

\bibitem{suddenferlaz}  L.~Fern\'andez-Jambrina and R.~Lazkoz,
 Phys.\ Rev.\  D {\bf 70}, 121503 (2004)
 [arXiv:gr-qc/0410124].

\bibitem{freeze} M.~Bouhmadi-L\'opez, P.~F.~Gonzalez-D\'\i az and
P.~Mart\'\i n-Moruno,
 \emph{Phys.\ Lett.\  B} {\bf 659}, 1 (2008).

 

\bibitem{tipler} F.J. Tipler, {Phys. Lett.} \textbf{A64}, 8 (1977). 
\bibitem{krolak} A. Kr\'olak, Class. Quant. Grav. \textbf{3},
267 (1986). 
\bibitem{puiseux}
 L.~Fern\'andez-Jambrina and R.~Lazkoz,
 Phys.\ Rev.\  D {\bf 74}, 064030 (2006)
 [arXiv:gr-qc/0607073].


\bibitem{tsagas} J.D. Barrow, C.G. Tsagas, 
\textit{Class.\ Quant.\ Grav.}\  {\bf 22}, 1563 (2005).


\bibitem{wsing} M.P. D\c abrowski, T. Denkiewicz, \emph{Phys. Rev. D} 
\textbf{79}, 063521 (2009).

\bibitem{loitering} Y. Shtanov, V. Sahni, \emph{Phys. Rev. D} 
 \textbf{71}, 084018 (2005).

\bibitem{barotrope} L. Fern\'andez-Jambrina, \emph{Phys. Rev. D} 
\textbf{82}, 124004 (2010).


\bibitem{HE} S.W. Hawking, G.F.R. Ellis,
\textit{The Large Scale Structure of Space-time}, Cambridge University
Press, Cambridge, (1973).

\bibitem{carlson} 
M.V. Fischetti, J.B. Hartle, B.L. Hu, \emph{Physical Review D}
\textbf{20}, 1757, (1979); A.A. Starobinsky, \emph{Physics Letters B} \textbf{91}, 99 (1980); T.
Azuma, S. Wada, \emph{Prog.  Theor.  Phys} \textbf{75}, 845 (1986); J. Haro, J. 
Amor\'os, E. Elizalde,
\emph{Physical Review D} \textbf{83}, 123528 (2011); E.D. Carlson,
P.R. Anderson, J.R. Einhorn, B. Hicks, A.J. Lundeen, [arXiv:
1607.01699]

\bibitem{vidotto} P. Singh, \emph{Class. Quantum Grav.} 
\textbf{26}, 125005 (2009); A. Corichi, P. Singh, \emph{Phys.  Rev.
D} \textbf{80}, 044024 (2009); P. Singh, F. Vidotto, \emph{Phys.  Rev.
D} \textbf{83}, 064027 (2011); K. Bamba, J. Haro, S.D. Odintsov, 
\emph{JCAP} \textbf{2013}, 008 (2013)

\bibitem{ellis} G.F.R. Ellis, B.G. Schmidt, {Gen. Rel. Grav.}
\textbf{8}, 915 (1977).

\bibitem{clarke} C.J.S. Clarke and A. Kr\'olak, Journ. Geom. Phys.
\textbf{2}, 127 (1985). 








 





\end{thebibliography}
\end{document}